\documentclass[12pt,a4paper]{article}
\usepackage{color,graphics,amsmath,epsfig,rotating}

\begin{document}

\setlength{\unitlength}{1mm}
\renewcommand{\arraystretch}{1.4}

%------------------------------------------------------------------------
% new definitions, abreviations, etc
%------------------------------------------------------------------------

\def\micromegas      {{\tt micrOMEGAs}}
\def\cpsh            {{\tt CPsuperH}}
\def\micro{\tt micrOMEGAs}

\def\ma{M_A}
\def\ra{\rightarrow}
\def\mneut{m_{\tilde{\chi}^0_1}}
\def\mchi{m_{\tilde{\chi}^0_i}}
\def\mneutt{m_{\tilde{\chi}^0_2}}
\def\mneuth{m_{\tilde{\chi}^0_3}}
\def\mneutf{m_{\tilde{\chi}^0_4}}
\def\mchar{m_{\tilde{\chi}^+_1}}
\def\mchart{m_{\tilde{\chi}^+_2}}
\def\msel{m_{\tilde{e}_L}}
\def\mser{m_{\tilde{e}_R}}
\def\mslo{m_{\tilde{\tau}_1}}
\def\mslt{m_{\tilde{\tau}_2}}
\def\msul{m_{\tilde{u}_L}}
\def\msur{m_{\tilde{u}_R}}
\def\msdl{m_{\tilde{d}_L}}
\def\msdr{m_{\tilde{d}_R}}
\def\msto{m_{\tilde{t}_1}}
\def\mstt{m_{\tilde{t}_2}}
\def\msbo{m_{\tilde{b}_1}}
\def\msbt{m_{\tilde{b}_2}}
\def\sw{s_W}
\def\cw{c_W}
\def\ca{\cos\alpha}
\def\cb{\cos\beta}
\def\sa{\sin\alpha}
\def\sb{\sin\beta}
\def\tb{\tan\beta}
\def\ssi{\sigma^{SI}_{\chi N}}
\def\si{\sigma^{SI}}
\def\sip{\sigma^{SI}_{\chi p}}
\def\ssd{\sigma^{SD}_{\chi N}}
\def\sd{\sigma^{SD}}
\def\sdp{\sigma^{SD}_{\chi p}}
\def\sdn{\sigma^{SD}_{\chi n}}
\def\msl{M_{\tilde l}}
\def\msq{M_{\tilde q}}
\def\bsg{B(b\rightarrow s\gamma)}
\def\bsmu{B(B_s\rightarrow\mu^+\mu^-)}
\def\btau{R(B\rightarrow\tau\nu)}
\def\Omg{\Omega h^2}
\def\sip{\sigma^{SI}_{\chi p}}
\def\amu{\delta a_\mu}
\def\lsp{\tilde\chi^0_1}
\def\neuto{\tilde\chi^0_1}
\def\neuti{\tilde\chi^0_i}
\def\neutt{\tilde\chi^0_2}
\def\neuth{\tilde\chi^0_3}
\def\neutf{\tilde\chi^0_4}
\def\chargi{\tilde\chi^+_i}
\def\charg{\tilde\chi^+_1}
\def\chargt{\tilde\chi^+_2}
\def\gluino{\tilde{g}}
\def\ul{\tilde{u}_L}
\def\ur{\tilde{u}_R}
\def\stau{\tilde{\tau}}
\def\sl{\tilde{l}}
\def\sq{\tilde{q}}
\def\bone{B^1}
\def\lkp{\gamma^1}

% Def. fuer groesser-ungefaehr:
\newcommand{\gsim}{\;\raisebox{-0.9ex}           {$\textstyle\stackrel{\textstyle >}{\sim}$}\;}

%=======================================================================
% Title
%=======================================================================

\begin{flushright}
   \vspace*{-18mm}
   Date: \today
\end{flushright}
\vspace*{2mm}

\begin{center}

{\Large\bf Dark matter in UED : the role of the second KK level} \\[8mm]

{\large   G.~B\'elanger$^1$, M.~Kakizaki$^{1,2}$,  A.~Pukhov$^3$}\\[4mm]
{\it 1) LAPTH, Univ. de Savoie, CNRS, B.P.110,  F-74941 Annecy-le-Vieux, France\\
     2)  Institute for Theoretical Physics, Hamburg University
  Luruper Chaussee 149, 22761 Hamburg, Germany\\
     3) Skobeltsyn Inst. of Nuclear Physics, Moscow State Univ., Moscow 119992, Russia 
}\\[4mm]

\end{center}

\begin{abstract}
We perform a complete calculation of the relic abundance of the KK-photon LKP 
in the universal extra dimension model 
including all coannihilation channels and all resonances. 
We show that the production of level 2 particles which decay dominantly into SM particles
contribute significantly to coannihilation processes involving level 1  KK-leptons.
% In particular the production of  $\gamma^2$ in coannihilation channels involving  increases the effective annihilation cross section and pushes the
As a result the preferred dark matter scale is increased to $R^{-1}=1.3$~TeV.
A dark matter candidate at or below the TeV scale can only be found in the non-minimal model by reducing the mass splittings 
between the KK-particles and the LKP. The LKP nucleon scattering cross section  is 
typically small, $\sigma < 10^{-10}$~pb, unless the KK-quarks are nearly degenerate with the LKP.
\end{abstract}

\section{Introduction}

One of the most attractive explanations to the dark matter(DM) problem is a new 
weakly interacting particle (WIMP) present in extensions of the standard model. 
Supersymmetry and extra dimension models are the leading candidates for physics 
 beyond the standard model that also propose a WIMP dark matter candidate. Among
 the extra dimension models, the UED scenario~\cite{Appelquist:2000nn}
   where all standard model particles are allowed to propagate freely in the bulk
    is of particular interest. In this model momentum conservation in the extra 
    dimensions entails conservation of KK number.
    Orbifolding is required to obtain chiral
zero modes from bulk fermions, and breaks extra dimensional momentum
conservation.  However, there remains a discrete subgroup, KK parity,
thus the lightest KK-odd particle is stable.
    In the minimal  universal extra dimension model
(MUED)  the dark matter candidate is in general a vector particle,
 $\bone$, the Kaluza-Klein (KK) level 1 partner of the U(1) gauge
boson. 
One characteristic feature of the model is that the dominant annihilation channels of $B^1$ are into lepton final states. 
%This has   UED models  % a prime
%candidate for a BSM interpretation of the excess in the positron  data published
The observations of  PAMELA~\cite{Adriani:2008zr} and Fermi~\cite{Abdo:2009zk} hinting 
at an excess  in the leptonic channel with no counterpart in the
antiproton channel has therefore renewed interest in the UED models.  Note however that 
this excess can be explained with astrophysical processes and
does not necessarily require a dark matter interpretation~\cite{Hooper:2008kg}.

In the MUED model  all KK states of a given level have nearly the same mass at tree-level, $n/R$, where $R$ is the size of the compact dimension.
The mass degeneracy is lifted only by standard model (SM) masses. Radiative corrections at the one-loop level further induced mass splittings among level 1
 particles, these mass splittings are however small for all weakly interacting particles.  
This means that co-annihilation channels naturally play an important role in the computation of the relic abundance
of dark matter. 
The relic abundance of $\bone$ in the MUED model was first computed  in ~\cite{Servant:2002aq} and 
included the coannihilation of SU(2) singlet KK leptons, the Next-to-Lightest KK particle (NLKP).
Apart from the mass difference entering the Boltzmann equation,  this calculation was performed   in the limit of degenerate masses for all particles at a given
KK level.  It was shown that  the effect of KK-leptons is to increase the dark matter relic
abundance. This is because the coannihilation channels are not as efficient as the main annihilation channels,
 the increase in the number of
effective degrees of freedom then induces an increase in the relic abundance  contrary to what usually occurs in the MSSM.
When the light Higgs mass is large (around 200GeV), the  KK-Higgs is the NLKP, coannihilation processes   
with  first KK level Higgs particles lead to a decrease of the relic abundance ~\cite{Matsumoto:2005uh} increasing the preferred mass scale for $B^1$ 
dark matter. 
 The relic abundance including all coannihilation channels as well as a precise evaluation of the KK masses  were later 
computed in ~\cite{Kong:2005hn,Burnell:2005hm}. The impact of the precise value of the mass splittings between the particles of the first KK level
and the LKP  was also analysed
by going beyond the MUED framework and treating the mass splittings as free parameters. Note that in  non minimal UED versions, it is possible to modify the
mass splittings by making different assumptions on the boundary terms at the cut-off scale.

At tree-level the second level KK particles have masses  twice as large as the ones of the first KK states. 
It is therefore natural to expect an enhancement of annihilation channels of 1st level KK particles due to the exchange of a s-channel 
2nd level KK particle near resonance.
For $\bone\bone$ annihilations, the only potential resonance is  the KK
partner of the Higgs, $h^2$. This particle either decays into $\bone\bone$ or into standard model particles.
It was shown in ~\cite{Kakizaki:2005uy} that the loop-induced decays into standard particles dominate. 
The  inclusion of $h^2$ in the annihilation processes thus reduces significantly the relic abundance~\cite{Kakizaki:2005uy,Kakizaki:2005en}.
Other resonance effects  occur in coannihilation channels, for example  $e^1 e^1\rightarrow Z^2$. The computation of the relic abundance including all 
coannihilation channels and level 2 resonances from the Higgs and gauge boson sector was performed in ~\cite{Kakizaki:2006dz}. 
The effect of level 2 fermions was  not included as those particles do not decay dominantly into SM particles. 
There is however an additional effect from level 2 particles that has been overlooked until now, it is the production 
of a KK-even level 2 particle in association with a SM particle in the final state.
This level 2 particle then decays through loop induced processes into standard model particles.
For coannihilation processes where channels with $B^2$ in the final state are usually kinematically accessible, 
there can be strong enhancement due to resonance effects from a variety of level 2 particles,  
for example the coannihilation process  $e^1 B^1 \rightarrow e^2 \rightarrow e B^2$ can be strongly enhanced by the contribution of the level 2
lepton. 
%When the production of a level 2 KK particle and a SM particle is kinematically open, 
%the coannihilation cross sections can be strongly enhanced. This is  especially the case  for coannihilation processes
%where the lightest level 2 particle, $B^2$ can be easily produced. 
The coannihilation processes can even dominate over the annihilation channels 
thus reducing the relic density of $\bone$.

In this paper we compute the relic abundance of the LKP 
including all coannihilation channels, all possible level 2 resonances as well as level 2
particles in the final state. 
We take into account electroweak symmetry breaking effects that were neglected in previous calculations. 
These will for example impact the masses of the KK-Higgs particles.
We consider the case where
the DM candidate is the KK partner of the photon, $\gamma^1$, rather than $B^1$.
Note however that the LKP is dominantly compose of $B^1$ with only a small $W^1$ component. 
The new contributions from annihilation channels into gauge bosons will be suppressed due to this small mixing angle. 
Finally we include couplings between level-2 gauge or Higgs bosons and standard model fermions. 
In particular we perform a  one-loop computation of the couplings of the level 2 KK-Higgs into quark pairs updating previous calculations
~\cite{Kakizaki:2006dz} 
by including the contributions from $n=1$ scalars and weak gauge bosons. 
We work  first in the context of a minimal UED model\footnote{Note that our version of the minimal UED 
model differs slightly from the MUED model of Ref.~\cite{Cheng:2002iz}, in particular  as concerns  the masses of  the  $n=2$ gauge bosons 
and the couplings of level 2 Higgses to SM fermions as detailed in section 2.}
 then in an extension of the minimal model by  
 treating the mass corrections in the fermion sector as arbitrary parameters. 
 This is done to illustrate  the strong dependence of the relic density on  mass corrections 
 to the KK states. Indeed the mass corrections influence the  
 coannihilation suppression factor, determine 
how strongly the  channels with level 2 KK-particles in the final state are kinematically suppressed
as well as whether the (co)-annihilation process benefits from a resonance enhancement.
 In the MUED model we find that 
the mass splitting is such that the second KK level in the final state 
play an important role. Thus the relic abundance is reduced and the 
preferred value for the DM mass, consistent with cosmological measurements, $\Omega h^2=0.1120\pm 0.0056$~\cite{Komatsu:2010fb},
is  shifted above the TeV scale. We also give predictions for direct detection rates and for KK particles production at the LHC.

The paper is organized as follows: in section 2 we briefly present the UED model and give explicit expressions 
for mass splitting of
level 2 particles and loop induced vertices between level 2 and SM particles. In section 3 we discuss DM (co-)annihilation channels and 
DM observables.   In section 4 the results for the relic
abundance and the detection rates are discussed. Section 6 contains our conclusions.

\section{The UED model}

 In the UED model all SM fields are upgraded to higher dimensional fields that
propagate in the bulk of the flat and finite spatial extra dimensions.
Among many space-time structures that can reproduce the SM as a low
energy effective theory, we focus on the simplest construction: the
five-dimensional space-time with the extra dimension compactified on
$S_1/Z_2$.  Integrating out the extra dimension leads for every 5D
field to an infinite tower of 4D Kaluza-Klein modes.  In the limit of
5D Lorentz invariance, the mass spectrum of the $n^{th}$ KK modes is
$n/R$, where R is the radius of the compactified dimension.  The $S_1$
compactification  destroys the 5D Lorentz invariance, and the
$Z_2$ orbifolding violates 5D momentum conservation.  The dispersion
relations of 5D particles are modified, resulting in shifts of the KK
mass spectrum from $n/R$ and in new decay patterns.
  The $Z_2$
orbifolding leads to chiral zero modes (the SM fermions) and removal
of the zero modes of the extra dimensional components of the gauge bosons,
and allows for 4D brane
interactions on the orbifold fixed points.  Even though the $Z_2$
orbifolding violates momentum conservation in the extra dimension, we
can retain a parity symmetry under the flip of the fifth coordinate.
From the four-dimensional viewpoint, this $Z_2$ parity prevents mixing
between even and odd KK levels, thus guaranteeing the stability of the
lightest KK-odd particle, the LKP.  The coupling of the KK modes are
generated from the 5D Lagrangian after expansion over the KK modes.

We first construct a gauge invariant 5D SM Lagrangian consistent with
the above observations.  Kinetic terms violating the 5D Lorentz
symmetry are compatible with the 5D gauge invariance, and can
practically produce the loop-induced mass shifts of the MUED model, as
we will see shortly.  We consistently implement electroweak symmetry
breaking: the vacuum expectation value of the Higgs scalar field is
involved not only in the KK electroweak gauge boson and KK fermion
mass matrices, but also in the KK Higgs boson mass splitting.
In addition, mixing angles caused by electroweak symmetry breaking
are taken into account in any interaction term.  Gauge-fixing functions
and ghost terms are introduced using 5D Goldstone and ghost fields.
As far as the bulk 5D Lagrangian  is concerned, the KK number is conserved
at each vertex.  Then, for phenomenological studies, we adjust the KK
mass shifts to the MUED ones, which are radiatively generated, and add
to our UED model KK-number-violating direct couplings between level 2
particles and SM particles as perturbation because such couplings are
necessarily induced at the loop level.  The complete description of
the Lagrangian and of the gauge fixing procedure is provided in
another publication~\cite{Kakizaki_uedmodel}.  Here we only describe the mass
spectrum including radiative corrections as well as loop induced
decays of level 2 particles into pairs of SM particles.

%%%%%

\subsection{Mass corrections}
The UED model is an effective field theory valid up to a cutoff scale $\Lambda$.  At loop level mass shifts are 
introduced by bulk corrections and by localized brane terms. 
Bulk corrections  affect only the gauge bosons KK states while localized brane terms 
induce mass shifts for all particles. In the minimal UED model one chooses vanishing  boundary  terms at the
cutoff scale. Nevertheless after running down to  the electroweak scale 
radiatively-generated bulk and brane terms  affect the mass spectrum lifting the degeneracy of
the KK levels. The mass corrections to  $B^n$ and $W^n$  were computed in Ref.~\cite{Cheng:2002iz}.
The squared mass matrix for neutral gauge bosons reads

\begin{eqnarray}
  {\cal M}_V^{2(n)} = 
    \left(
    \begin{array}{cc}
      (n/R)^2 +  m_Z^2 c_W^2 +\delta m^2_{W^n}& - m_Z^2 c_W s_W \\
       - m_Z^2 c_W s_W  &   (n/R)^2  + m_Z^2 s_W^2+ \delta m^2_{B^n}
    \end{array}
  \right)\, 
  \label{eq:mv}
  \end{eqnarray}
where $s_W,c_W$ are the sine and cosine of  the weak mixing angle and 

\begin{equation}
\delta m^2_{W^n} = -\frac{5}{2} \frac{g^2 \zeta(3)}{16\pi^4} \frac{1}{R^2} 
 + \frac{15 g^2}{2}\frac{n^2}{R^2}\frac{1}{16\pi^2} \log\left(\frac{\Lambda^2}{\mu^2}\right)
\label{eq:mw2}
\end{equation}

\begin{equation}
\delta m^2_{B^n} = -\frac{39}{2} \frac{g'^2 \zeta(3)}{16\pi^4} \frac{1}{R^2} 
 - \frac{g'^2}{6}\frac{n^2}{R^2}\frac{1}{16\pi^2} \log\left(\frac{\Lambda^2}{\mu^2}\right)
 \label{eq:mb2}
\end{equation}
The renormalization scale $\mu$ is chosen to be the mass scale of the 1$^{st}$ KK mode, $\mu = R^{-1}$,
and 
$g,g'$ are the SU(2) and U(1) gauge couplings respectively. The mass of the LKP, $\gamma^1$, is obtained after 
diagonalisation of the $B^1, W^1$ mass matrix.
\begin{eqnarray}
  A_\mu^{(n)} & = & W_\mu^{3(n)} \sin \theta_W^{(n)} 
  + B_\mu^{(n)} \cos \theta_W^{(n)}\, ,
\nonumber \\
  Z_\mu^{3(n)} & = & W_\mu^{3(n)} \cos \theta_W^{(n)} 
  - B_\mu^{(n)} \sin \theta_W^{(n)}\, ,
\end{eqnarray}
where $\theta_W^{(n)}$ is the $n^{th}$ level mixing angle which  is near 0 for $n\geq 1$. The mass of the LKP is given approximately by 
the mass of $B^1$ in Eq.~\ref{eq:mv} and deviates from $1/R$ at the permil level or less. 
To take into account the loop corrections to the mass terms in a gauge invariant manner in  the 5D Lagrangian we 
introduce a correction to the fifth components of the  SU(2) and U(1) gauge fields, namely,
\begin{equation}
{\cal L}^{5D} = -\frac{1}{4} B_{\mu\nu} B^{\mu\nu} + \frac{Z_B}{2} B_{\mu 5} B_5^{\mu} +\frac{1}{4} {\bf W}_{\mu\nu} {\bf W}^{\mu\nu} + 
\frac{Z_W}{2} {\bf W}_{\mu 5} {\bf W}_5^{\mu}
\end{equation}

This improved tree-level Lagrangian reproduces the loop-corrected mass terms once the parameters  
 $Z_B,Z_W$ are fixed so that the new mass matrix matches Eq.~\ref{eq:mv}, see ref.~\cite{Kakizaki_uedmodel} for a complete description.
 The gauge boson  mass matrix now reads
\begin{eqnarray}
 {\cal M}_V^{2(n)} =\left(
    \begin{array}{cc}
      Z_W (n/R)^2 + m_Z^2 c_W^2 & - m_Z^2 c_W s_W \\
       - m_Z^2 c_W s_W  &  Z_B (n/R)^2 + m_Z^2 s_W^2 
    \end{array}
  \right)\, 
\end{eqnarray}

As long as $m_h$ is light (near 120GeV), the next to lightest KK particles are the right-handed KK leptons whose mass corrections
are also governed by the U(1) coupling. The mass eigenstates are obtained after diagonalisation of the KK lepton mass matrix,
\begin{equation}
  {\cal M}^{(n)}_l =
  \left( \begin{array}{cc}
   n/R +\delta m_{l^n_L}&  m_l\\
  m_l&   -n/R -\delta m_{l^n_R}
  \end{array}\right)  
  = \left( \begin{array}{cc}
 Z_{l_L}  n/R &  m_l\\
  m_l&   -Z_{l_R} n/R 
  \end{array}\right) 
  \label{eq:ml}
\end{equation}
where  the  parameters $Z_{l_L},Z_{l_R}$  describe the loop-improved masses in the tree-level gauge invariant 5D Lagrangian, explicitly 
\begin{equation}
{\cal L}^{5D} = \overline{\psi}_L i \gamma^\mu D_\mu \psi_L - Z_{l_L} \overline{\psi}_L  \gamma^5 D_5 \psi_L
+ \overline{\psi}_R i \gamma^\mu D_\mu \psi_R - Z_{l_R} \overline{\psi}_R  \gamma^5 D_5 \psi_R
\label{eq:zl}
\end{equation}
where $D_{\mu,5}$ are covariant derivatives. In the MUED model, the corrections for the KK partners of singlet leptons read
\begin{equation}
\delta m_{l^n_R}= \frac{9}{4} \frac{g'^2}{16\pi^2} \frac{n}{R} \log\left(\frac{\Lambda^2}{\mu^2}\right).
\end{equation} 
These corrections give a mass splitting at the  percent level between the LKP and the NLKP. 
The mass splittings for the KK lepton doublets and the KK gauge bosons are governed by the SU(2) coupling and are at the
few percent level while those for
coloured particles are governed by the SU(3) coupling. 
For the top quark an additional mass splitting is induced by a term proportionnal to
the Yukawa coupling.  The mass difference between  the coloured particles and the LKP are generally large enough (15-20\%) that 
 the Boltzmann  factor suppresses  the
contribution of the coloured states in coannihilation channels.
To implement the mass corrections for KK-quarks we introduce three new parameters in the Lagrangian $Z_{Q_L},Z_{d_R},Z_{u_R}$ and follow the same
prescription as for leptons.

In  the Higgs sector,  the KK charged Higgs $h^{\pm n}$ is the lightest
particle, the CP-odd Higgs $a^n$ the next to lightest, and the CP-even
Higgs $h^{n}$ the heaviest at each KK level, the masses read
\begin{eqnarray}
m^2_{h^n} =\frac{n^2}{R^2}+ m_h^2 +\delta m^2_{H^n}\nonumber\\
m^2_{a^n} =\frac{n^2}{R^2}+ m_Z^2 +\delta m^2_{H^n}\nonumber\\
m^2_{h^{\pm n}} =\frac{n^2}{R^2}+ m_W^2 +\delta m^2_{H^n}\nonumber\\
\end{eqnarray}
where
\begin{equation}
\delta m^2_{H^n} = \left( \frac{3}{2}g^2 +\frac{3}{4} g'^2-\lambda_h \right)\frac{n^2}{R^2}\frac{1}{16\pi^2} 
\log\left(\frac{\Lambda^2}{\mu^2}\right)
\label{eq:dmh}
\end{equation}
where the Higgs quartic coupling, $\lambda_h$  is related to the mass $m_h=\sqrt{\lambda_h} v$. 
Since  the quartic coupling induces a negative mass correction to
the Higgses, for a large value of $m_h$ (and $\lambda_h$) the charged Higgs can become the NLKP or 
even the LKP ~\cite{Matsumoto:2005uh}.
In the 5D Lagrangian,  corrections to the KK masses of the Higgs doublet $\phi$ are
 taken into account by introducing the parameter 
$Z_\phi$ 
\begin{equation}
{\cal L}^{5D} = D^\mu\phi^\dagger D_\mu\phi - Z_\phi D_5\phi^\dagger D_5\phi -\mu^2\phi^\dagger\phi
\label{eq:zh}
\end{equation}
The complete mass corrections in the MUED model are given in Ref.~\cite{Cheng:2002iz}.
Note that when computing the spectrum we fix the gauge couplings at the electroweak scale.

\subsection{Decays of level 2 particles}
\label{sec:decay}

At tree-level the $n=2$ KK-particles can decay either into two $n=1$ KK-particles or into another $n=2$ KK particle and a SM one.   
Both these channels are kinematically suppressed so that the dominant decay mode can be a loop-induced two-body process into SM particles. 
In particular the $n=2$ partner of the LKP will decay dominantly into SM fermions.  
The coupling of $\gamma^2$ to standard model fermions is a loop-induced process involving triangle and self-energy diagrams 
with  level 1 fermions and a level 1  gauge boson.  The vertices $B^2 \bar{f} f$ and $W^2 \bar{f} f$ have been computed 
in ~\cite{Cheng:2002iz} and are  related to the mass corrections from the boundary terms.
The dominant loop-induced coupling comes from the level 1 gluon and level 1 quark 
exchange diagram in the $B^2 q\bar{q}$ vertex. 
In particular  the vertex $B^2 t\bar{t}$ receives corrections that are proportionnal to the top Yukawa, $y_t$. 
The loop-corrected vertex reads
\begin{eqnarray}
{\cal L} &=& -\frac{g_1}{\sqrt{2}} \frac{1}{32\pi^2}\log\left( \frac{\Lambda}{\mu} \right)
\bar{t} \gamma^\mu 
\left( Y_{t_L} (\frac{7}{24} g'^2+\frac{27}{8}g^2+6 g_s^2-\frac{3}{2}y_t^2) (1-\gamma_5)\right.\nonumber\\
&+&  \left. Y_{t_R} (\frac{13}{6} g'^2+6 g_s^2- 3 y_t^2) (1+\gamma_5)\right) t B^2_\mu
\end{eqnarray} 
where $Y_{t_L},Y_{t_R}$ are the hypercharge of $t_L$ and $t_R$.
Note that  there is a partial cancellation between the gluon and Yukawa contribution. 
For the light quarks one gets the same expression neglecting the Yukawa term 
while the couplings to leptons  are suppressed as they receive no contribution from gluon exchange nor from large Yukawas. 
The width of $\gamma^2$ is around 1 GeV when  $R^{-1}=1$~TeV and  
the branching ratios do not vary much with $R$ or $\Lambda R$, with 13.9\% into $t\bar{t}$ , 29.7\% into  $c\bar{c}$ and $u\bar{u}$, 
and 8.6\% into each of the d-type quarks.

The Higgs bosons, $h^2$, $a^2$ and $h^{2\pm}$, are the other level 2
particles where the dominant decay mode is into SM particles.  The
level 2 CP-even Higgs, $h^2$, predominantly decays into $t \bar{t}$
through the radiatively generated vertex $h^2 t \bar{t}$.  The
$t^1$-$g^1$ one-loop contribution to the vertex $h^2 t \bar{t}$ is
found in Ref.~\cite{Kakizaki:2006dz}.  In addition, we
include the vertex corrections from the $t^1$-$W^1$ and $t^1$-$B^1$
one-loop diagrams, as well as the contributions from the kinetic and
mass mixings between the level 2 particles and the SM particles on the
external legs.  The latter contributions stem not only from the gauge
interactions but also from the top-Yukawa interaction and Higgs
self-interaction, which allow the level 1 Higgs bosons to run in the
loop.  The full expression for the vertex $h^2 t \bar{t}$ at the
one-loop level is given by
\begin{eqnarray}
  {\cal L} = \frac{y_t}{96 \pi^2} \left( 16 g_s^2 - \frac{39}{4}g^2 
  + \frac{4}{3} g^{\prime 2} - 9 y_t^2 + 3 \lambda_h \right) 
  \log \left( \frac{\Lambda}{\mu} \right) h^2 \bar{t} t\, .  
\end{eqnarray}
The coefficients of the vertices, $a^2 t \bar{t}$ and $h^{2-} t
\bar{b}$, have the same form due to gauge invariance.  In our
analyses, we omit the loop-induced $h^2 h W$-type vertices as such
interactions are suppressed by the weak coupling constant.  Loop
diagrams that involve the Higgs vacuum expectation value are also
neglected.  Finally the loop induced vertices to lighter quarks, including $\overline{b}b$ are negligible. 
Given the above interactions, the total decay widths of
$h^2$ ($a^2$,$h^{2\pm}$) are 395(400,401)~MeV for $R^{-1}=1.3$ TeV, $m_h=120$~GeV and $\Lambda R=20$.  
The branching ratio of each level 2 Higgs boson
into the third generation SM quarks is more than $99$\%.

Notice that in the MUED model the kinetic and mass mixing terms, which
contribute to the direct couplings of level 2 particles with SM
particles, are proportional to the radiatively-induced mass shifts of
the level 2 particles.  In the scenarios with arbitrary mass
splittings we discuss later, we use the same KK-number-violating
interactions as in the MUED model.  As long  as the arbitrary mass
splittings are introduced at the tree level with the cutoff scale
fixed as assumed in this paper, only the argument of the logarithm in
the KK-number-violating couplings is affected.  Since we use the
leading log approximation, to take into account such effects is
meaningless.

\section{Relic abundance }

To compute the relic density of dark matter, we have implemented the
UED model in CalcHEP~\cite{Pukhov:2004ca} and \micromegas2.4~\cite{Belanger:2006is,Belanger:2010gh}.
For this we relied on LanHEP
and the new facilities to project the 5D Lagrangian into a 4D one
~\cite{Semenov:2010qt}.  Implementations of the MUED model in CalcHEP~\cite{Datta:2010us}  
and in FeynRules~\cite{Christensen:2009jx} are available and aim primarily at collider studies.
Here, we briefly describe differences between the implementation of
our phenomenological UED model and that of the MUED.  First, we have
implemented the classical 5D SM Lagrangian with the wave function
factors, $Z_i$, which allow for arbitrary mass splittings provided
that the mass corrections of KK particles are proportional to $n$.
The Higgs sector and the electroweak symmetry breaking are fully taken
into account: for example, the mass splitting among the KK Higgs
bosons, Eq.(9), is automatically generated.  For simplicity, we
neglected the SM quark mixing angles and the renormalization group
running of the coupling constants, both of which are included in
Ref.~\cite{Datta:2010us}. In the realisation of Ref.~\cite{Christensen:2009jx} the level 2 KK-particles
are not included as well as    the charged and CP-odd level 1 KK-Higgses.  
Notice that there exist KK Goldstone and ghost
fields corresponding to the higher level massive gauge bosons,  these were not considered in the
preceding implementations.  We have
implemented both the unitary and the Feynman gauges, and checked that the
5D gauge invariance is retained.  The loop induced vertices of the
level 2 KK-particles with SM particles described in
section~\ref{sec:decay} are then added to the model file by hand, so
that the decay widths of the level 2 particles are automatically
computed.  The model will be described in ~\cite{Kakizaki_uedmodel}.

To discuss MUED phenomenology, the wave function factors are adjusted
to reproduce the loop-induced mass spectrum of the level 1 KK
particles: the masses of the higher KK particles are given by integer
multiples of those of the level 1 KK particles.  In the MUED model,
this holds true for all corrections except the bulk corrections on the
gauge bosons, Eqs.~\ref{eq:mw2},\ref{eq:mb2}, and those induced by  a shift in the
renormalization scale.  In total, our approximation leads to a shift
on the mass of level 2 KK gauge particles less than per-mil for
$m_{B^2}$ and 1.2\% on the mass of $W^2,Z^2$ as compared to the MUED
case.  In comparison, changing the cut-off scale from $\Lambda R=20$
to $\Lambda R=50$ leads to 1.6\% shifts on these masses.  Note however that even a  small shift for
the mass of  a level 2 KK-particle  can give rise to a noticeable discrepancy. For example,  in
most of the parameter space of our model, $\gamma^1 h^{1+} \to
h^{2+} \to t\bar{b}$ gives the dominant contribution to the relic
abundance among loop-induced resonant processes, while $h^{2\pm}$ is
light enough that such a pole is not reached at the LKP decoupling temperature in
the MUED model, as we will see.

The relic abundance calculation when coannihilation channels are present involve generalizing to 
the effective annihilation cross section. To display explicitly the dependence on the mass differences we write $\langle \sigma v \rangle$ as
\begin{equation}
  \langle \sigma v \rangle =\frac{\displaystyle\sum_{i,j} \langle \sigma v\rangle_{ij} g_i g_j exp^{-(\Delta m_i+\Delta m_j)/T_f}}
  {\left(\displaystyle\sum_{i} g_i exp^{-\Delta m_i/T_f}\right)^2}
  \label{eq:coan}
\end{equation}
 where $\langle \sigma v\rangle_{ij}$ is the thermally average cross section for annihilation of any pair of $n=1$ KK particles into KK-even particles 
 and $g_i$ are the number of degrees of 
 freedom.  The Boltzmann suppression factor is $B_i=exp^{-\Delta m_i/T_f}$ where $T_f\approx m_{\rm LKP}/25$ and $\Delta m_i=m_i - m_{\rm LKP}$.

When computing the relic abundance we sum over all $n=1$ KK particles and include all possible SM particles in the final state 
as well as the KK-even $n=2$ particles. Indeed as mentionned above both $\gamma^2$ and the level 2 scalar, pseudoscalar and charged Higgs
($h^2,a^2,h^{2\pm}$) 
decay preferentially into SM particles. Furthermore their decay rate, typical of a weak interaction process, is much 
faster than the Hubble expansion rate at freeze-out temperatures. These processes therefore contribute to the rate of annihilation of dark matter particles. 
Other  $n=2$ KK particles, notably KK-leptons, can also be produced in 
the final state in coannihilation processes.  These particles can decay into other $n=1$ KK states 
as well as into SM particles.   One can show that the contribution of $n=2$ KK fermions in the final state  to the  
effective annihilation cross section can be taken into account by multiplying their production cross section by the branching fraction 
of KK fermions into SM particles. 
We have modified \micromegas~ to take this factor into account. 
Note however that the contribution of KK leptons in the final state is only at the percent level so this effect is small. 
The most general case with  large branching fractions of heavier level 2 particles into a lighter level 2 particle and
a SM particle would necessitate modification of the Boltzmann equation for the computation of the relic abundance. However such cases are not relevant in
the UED model under consideration.

We have also checked the  consistency of our numerical results for the main annihilation channels with Ref.~\cite{Kong:2005hn},
good agreement  was found when we removed the contribution of the level 2 particles coupling to SM particles. 
Furthermore our results qualitatively agree with ~\cite{Kakizaki:2005uy} 
when we ignore level 2 particles in the final states 
(but include $h^2$ exchange). Some numerical differences are found due to 
the more accurate computation of the $h^2$ decay width as described in section~\ref{sec:decay}. 
%
%Furthermore our results agree with ~\cite{Kakizaki:2005uy} when we ignored  level 2 particles in the final states (but included $h^2$ exchange),
%a difference of the order of one  percent between the two results is due to the more accurate computation of the 
% $h^2$ decay width as described in section~\ref{sec:decay}. 

The relic abundance calculation  can be rather slow in \micromegas~ because of the very large number of 
coannihilation channels of the UED model.
To avoid computing unnecessary coannihilation channels in the calculation of the contribution of individual channels to the relic
abundance we do not include channels for  which the Botlzmann suppression factors $B_i B_j<0.01$ 
(this is done by setting $B_{eps}=0.01$ in \micromegas). In the computation of the relic abundance on the other hand
we set  $B_i B_j<0.00001$. Note however that  setting $B_{eps}=0.01$ is sufficient to  compute  $\Omega h^2$  
with an accuracy below 1\% as we have tested with several input parameters.

\section{Results} 

The minimal UED model contains only three free parameters: $R^{-1}$, $\Lambda$ and $m_h$. We vary  the mass scale of KK states 
in the range 0.4-1.8~TeV. Indirect constraints from electroweak precision observables~\cite{Appelquist:2002wb,Gogoladze:2006br} 
 and  rare processes  such as  $b\rightarrow s\gamma$~\cite{Agashe:2001xt,Haisch:2007vb} set the
  lower bound between $R^{-1}>0.4-0.6$~TeV
 while larger values will lead to too much dark matter. 
 We vary $m_h=114.4-300$~GeV to comply with the LEP limit~\cite{Nakamura:2010zzi} and to include the whole region 
where the LKP is a neutral particle. For the cut-off scale we choose the range
$\Lambda R=20-50$. The upper range for the cut-off scale is
motivated by the computation in Ref.~\cite{Bhattacharyya:2006ym} of the renormalization group running of the couplings  which shows that 
the U(1) coupling blows up at $E=50$~TeV when  $R^{-1}=1$~TeV.

\begin{figure}[!ht]
\setlength{\unitlength}{1mm}
\includegraphics[height=6.1cm]{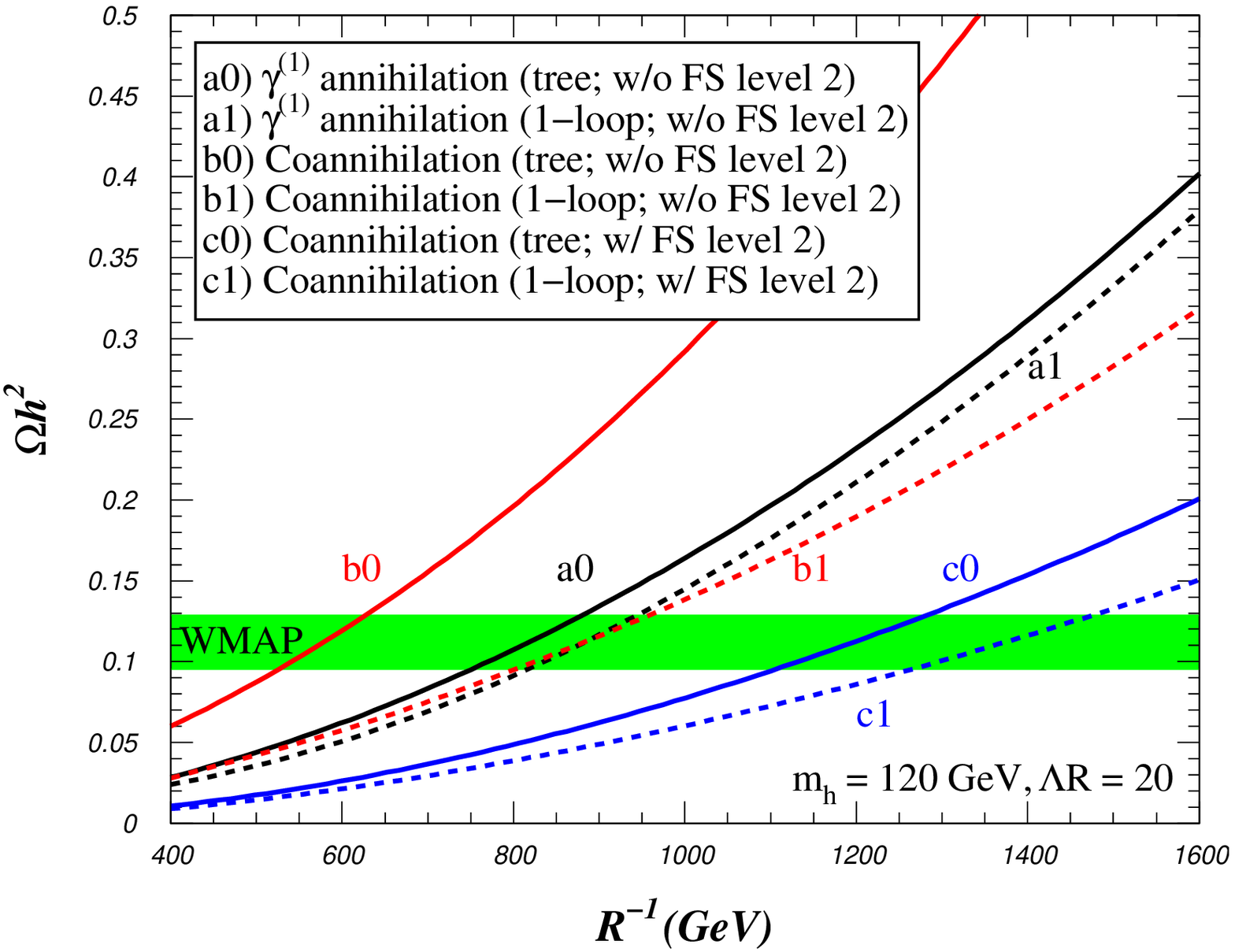}\hspace*{-1mm}
\includegraphics[height=6.1cm]{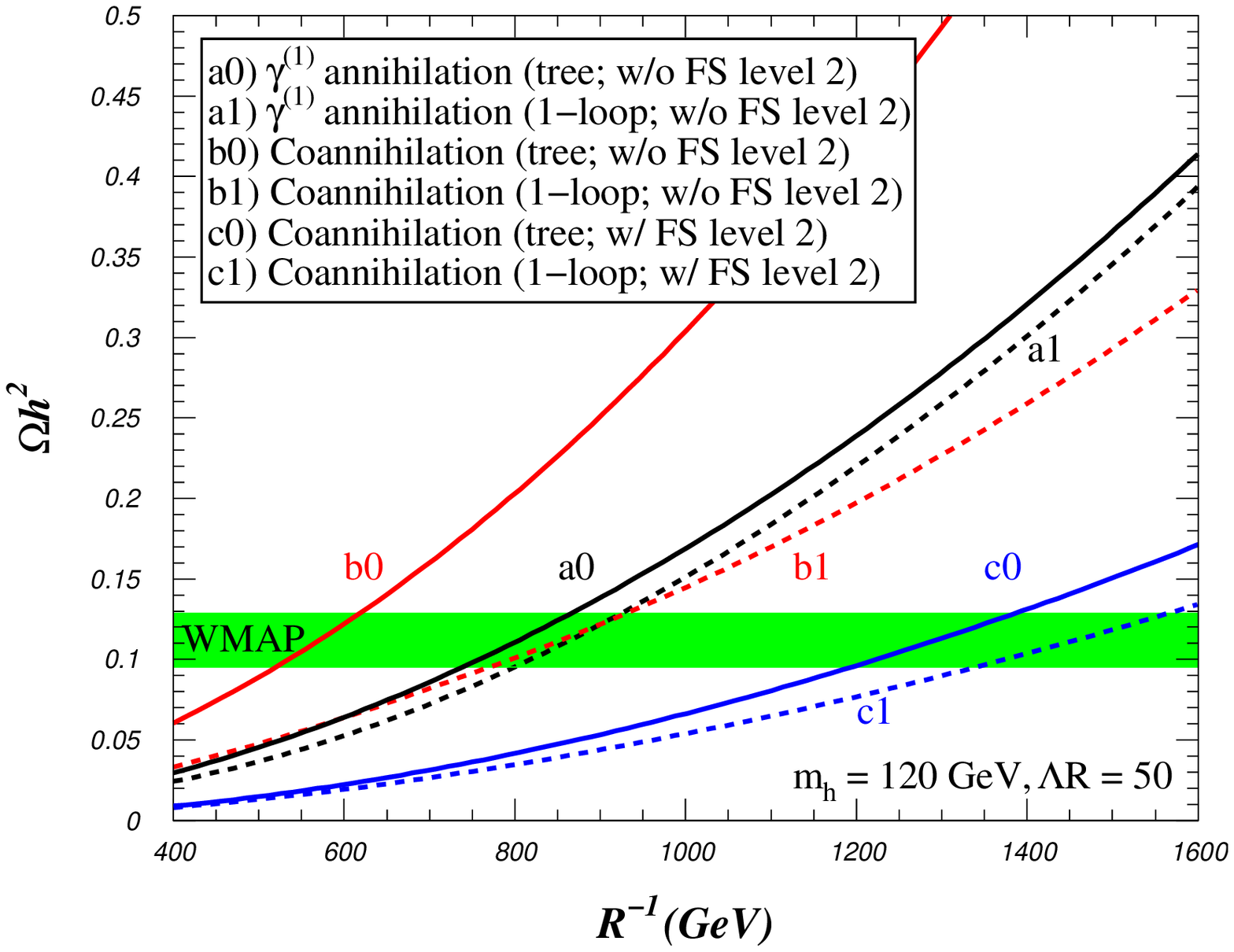}
\vspace{-.02cm}
\caption{$\Omega h^2$ as function of $R^{-1}$ for $m_h=120$~GeV, $\Lambda R=20$ (left) and $\Lambda R=50$ (right)
including different processes as specified on the figure. Here 1-loop stands for 
one-loop couplings between level 2 and SM particles. The shaded region  corresponds to the $3\sigma$ preferred region
obtained by WMAP~\cite{Komatsu:2010fb}.} 
\label{fig:omega}
\end{figure}

For $m_h=120$~GeV, $\Lambda R=20$, 
the value of  $\Omega h^2$ computed with only the main annihilation channels shows that $\gamma^1 \gamma^1$ annihilate
efficiently and that a rather heavy mass  is preferred ($m_\lkp=800-950$~GeV), see Fig.~\ref{fig:omega}.
The dominant channels do not vary much with the scale, for $R^{-1}=1$~TeV they are into $l \bar{l}$ (19\% each), $\bar{t}t$,
(21\%) $\bar{q}{q}$ (17\% altogether) and $\nu\bar\nu$ (3\%). 
The  contribution from W,Z boson final states is  around $\approx 1\%$ and is  
suppressed due to the very small  $B^1-W^1$ mixing. The enhancement of the $t\bar{t}$ channel as compared to other quarks is due to
the contribution of $h^2$. In Fig.~\ref{fig:omega}, we also display the value of $\Omega h^2$ when
neglecting the $h^2$ exchange in s-channel, this is done by removing the loop-induced vertex $h_2 f\bar{f}$.  
The contribution of $h^2$ induces   roughly 
a  10\% decrease in $\Omega h^2$ as was found previously ~\cite{Kakizaki:2005uy}.  

Including  all coannihilation channels while forbidding the production of $\gamma^2$ and other level 2 particles as well as loop-induced couplings between level
2 particles and SM ones 
leads to an increase in the relic density ~\cite{Servant:2002aq}. 
This is because  the coannihilation
cross sections (predominantly those involving the KK partners of singlet leptons) are typically weaker than the ones for  
$\gamma^1\gamma^1$ 
annihilation. Thus the contribution of coannihilation channels, despite that they do not suffer much from a Boltzmann suppression factor, 
$B_i \approx 0.9$,  is more than compensated by the increase
in the effective number of degrees of freedom, see Eq.~\ref{eq:coan}.
This is in sharp contrast with supersymmetry where coannihilation generally decreases the relic density as the coannihilation 
channels typically have much larger cross section than the main channel. 
A large number of coannihilation processes contribute to the effective annihilation cross sections, primarily coannihilation with Higgs,
$\gamma^1 H^1 \rightarrow t, \bar{t} (\bar{b})$  as well as a host of other self annihilation processes of the type
$l^1 \bar{l}^1 \rightarrow X\overline{X}, H^1 H^1\rightarrow X\overline{X'}, l^1 H^1\rightarrow X\overline{X'}, l^1 V^1\rightarrow X\bar{X'}$, 
each individual channel contributing to a small fraction (less than 1\%) of the total effective annihilation cross  section.  
Here $l$ stands for $e,\mu,\tau$, $V^1$ stands for $W^1,Z^1$ and $H^1$ for any of the neutral or charged Higgs. Coannihilations processes involving quarks are completely
negligible.
Adding the loop-induced couplings between level 2 KK particles  and SM particles has a significant impact on the 
relic density, see Fig.~\ref{fig:omega}. 
This is mainly because the new contribution from the process $\gamma^1 h^{1+} \to
h^{2+} \to t\bar{b}$ benefits from a resonance enhancement thus increasing significantly the effective annihilation 
cross section. 
This result depends very sensitively on the mass of the level-2 particle, a small downward shift in the mass, 
such as in the MUED model used in ~\cite{Kakizaki:2006dz}, where
the renormalization scale is set to $\mu = 2 R^{-1}$ for the level 2 masses,
means that the pole effect is avoided at the LKP decoupling temperature.  When including the contribution of 
$h^2$ and neglecting level 2 KK-particles
in the final state,   the prediction for the relic abundance is close to the one obtained
including  only annihilation processes.

\begin{figure}[!ht]
\setlength{\unitlength}{1mm} \centerline{\epsfig{file=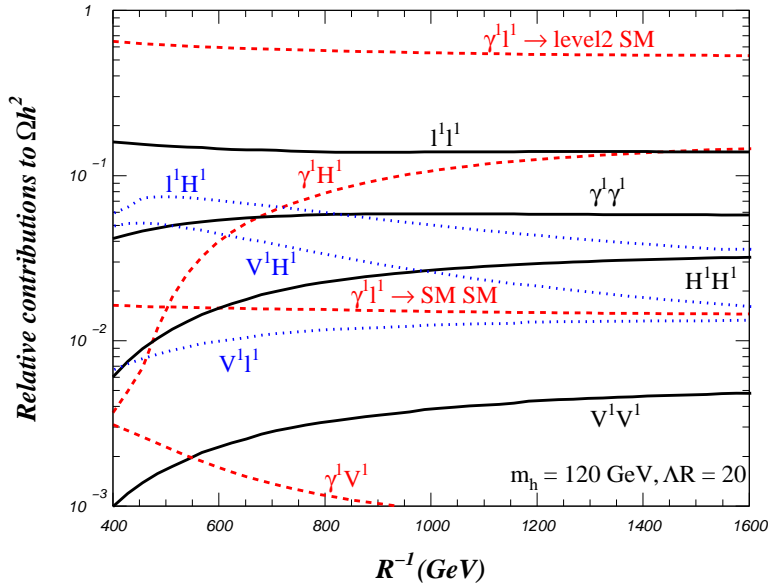,width=10cm,angle=0}}
\vspace{-.1cm}
   \caption{Relative contributions to the relic abundance at the LKP decoupling temperature as a
      function of $R^{-1}$ for $m_h=120$ GeV, $\Lambda R=20$.  Here,
      summation over a class of initial states and all possible final states is performed
      with the exception of $\gamma^1 l^1\rightarrow$ level 2 SM which includes only processes with one level 2 particle in the final state
      and  $\gamma^1 l^1 \rightarrow$ SM SM which includes only SM particles in the final state.
       $l^1$ stands for $e^1,\mu^1,\tau^1,\nu^1_i$, $V^1$ for $W^1,Z^1$ and $H^1$ for $a^1,h^{1\pm},h^1$. 
       All remaining channels contribute less than 1\%.}     
\label{fig:channel}
\end{figure}

When allowing level-2 particles in the final state, mainly $\gamma^2$ and $h^2,a^2,a^{\pm 2}$, the relic abundance 
decreases sharply shifting the preferred value of the DM mass above the TeV scale. 
This is due to the important contribution of the coannihilation channels ($l^1 \gamma^1\rightarrow l \gamma^2$)
that are enhanced by the exchange near resonance of the $n=2$  KK singlet lepton. 
Together these channels make up more than 50\% of the (co)annihilation channels.
As previously, other coannihilation channels each contribute to a small fraction of the total effective cross section.  
The contribution of the most important channels is illustrated in Fig.~2, where we have summed the contribution of all leptons in the initial states and
all SM particles in the final state. Coannihilation channels involving lepton pairs contribute around 15\% and their contribution 
is comparable to the one of Higgs channels $\gamma^1 H^1$ at large values of $R^{-1}$. Contributions 
of the order of a few percent
are found for the annihilation channels, $\gamma^1\gamma^1$,  as well as coannihilations of the type $l^1 H^1$, $H^1 H^1$ or 
$\gamma^1 l^1$ into only SM particles. This still leaves around 10\% contribution from all 
remaining channels, among these one finds 
notably channels involving gauge bosons such as $V^1 H^1$ or $V^1 l^1$.

The value of the cut-off scale $\Lambda$ has an impact on the mass of the KK particles through logarithmic one-loop corrections, Eq.~\ref{eq:dmh}.  
Increasing the scale to $\Lambda R=50$ leads
to heavier KK  particles, in particular for KK lepton doublets and  KK quarks, and has an impact on $\Omega h^2$. 
For example when ignoring the level 2 particles in the final state the contribution of coannihilation channels with KK leptons 
suffers from a larger Bolzmann suppression
factor, this is partly compensate by an increase in the contribution of the $h^{2+}$ pole (as well as the one of $h^2,a^2$) and therefore in the contribution of the coannihilation channels involving Higgs.
The net effect is an increase in $\Omega h^2$ by around 5\%. On the other hand in the complete calculation including all channels, 
both the  increase in the contribution of the $h^{2+}$ pole and the increase in the pole contribution of KK leptons which dominate  the effective cross section
lead to a 15\% decrease of $\Omega h^2$. This implies an additional  shift of close to 100GeV in the preferred range for the DM mass. 

\begin{figure}[!ht]
\setlength{\unitlength}{1mm} 
\includegraphics[height=6.3cm]{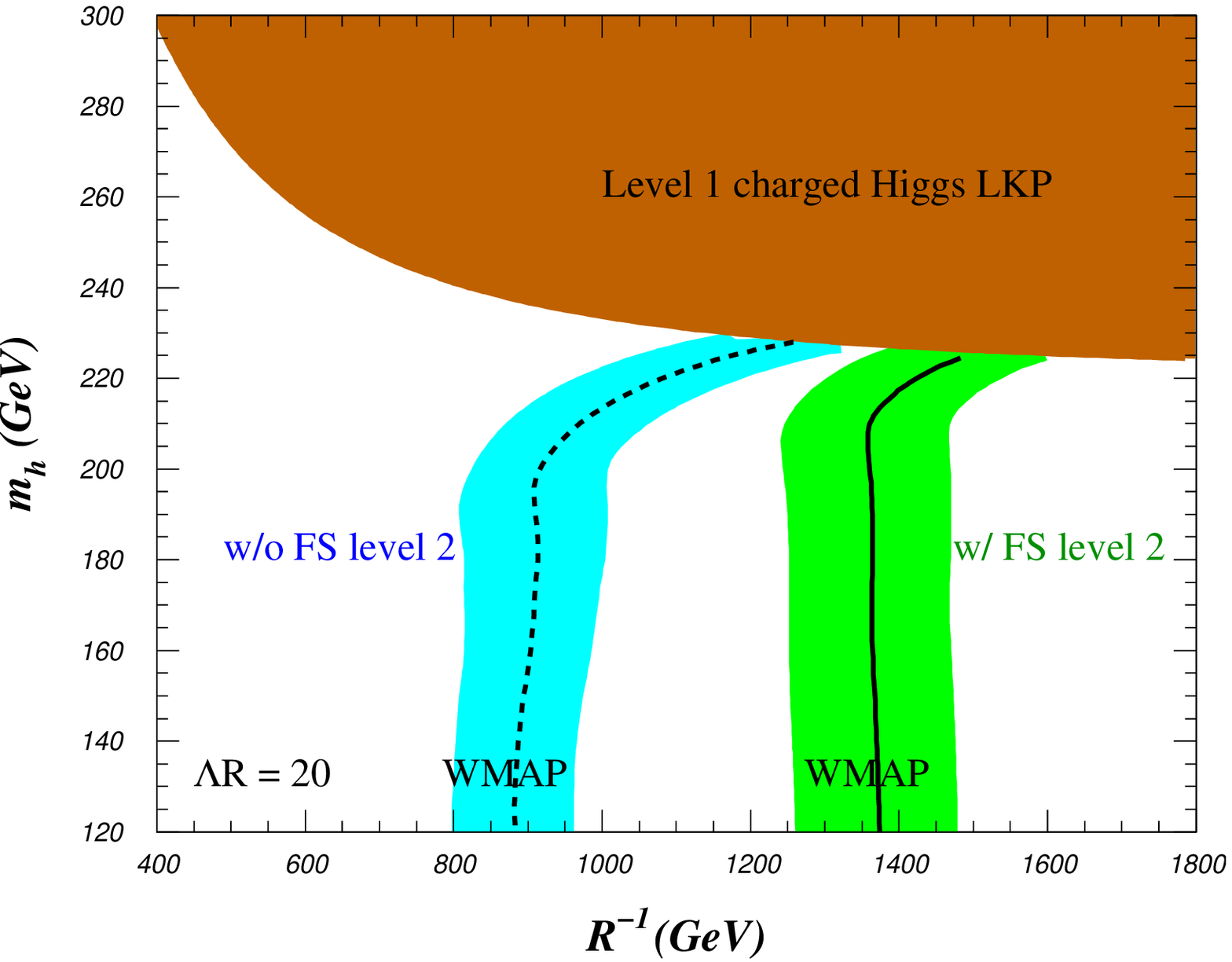}\hspace*{-1mm}
\includegraphics[height=6.3cm]{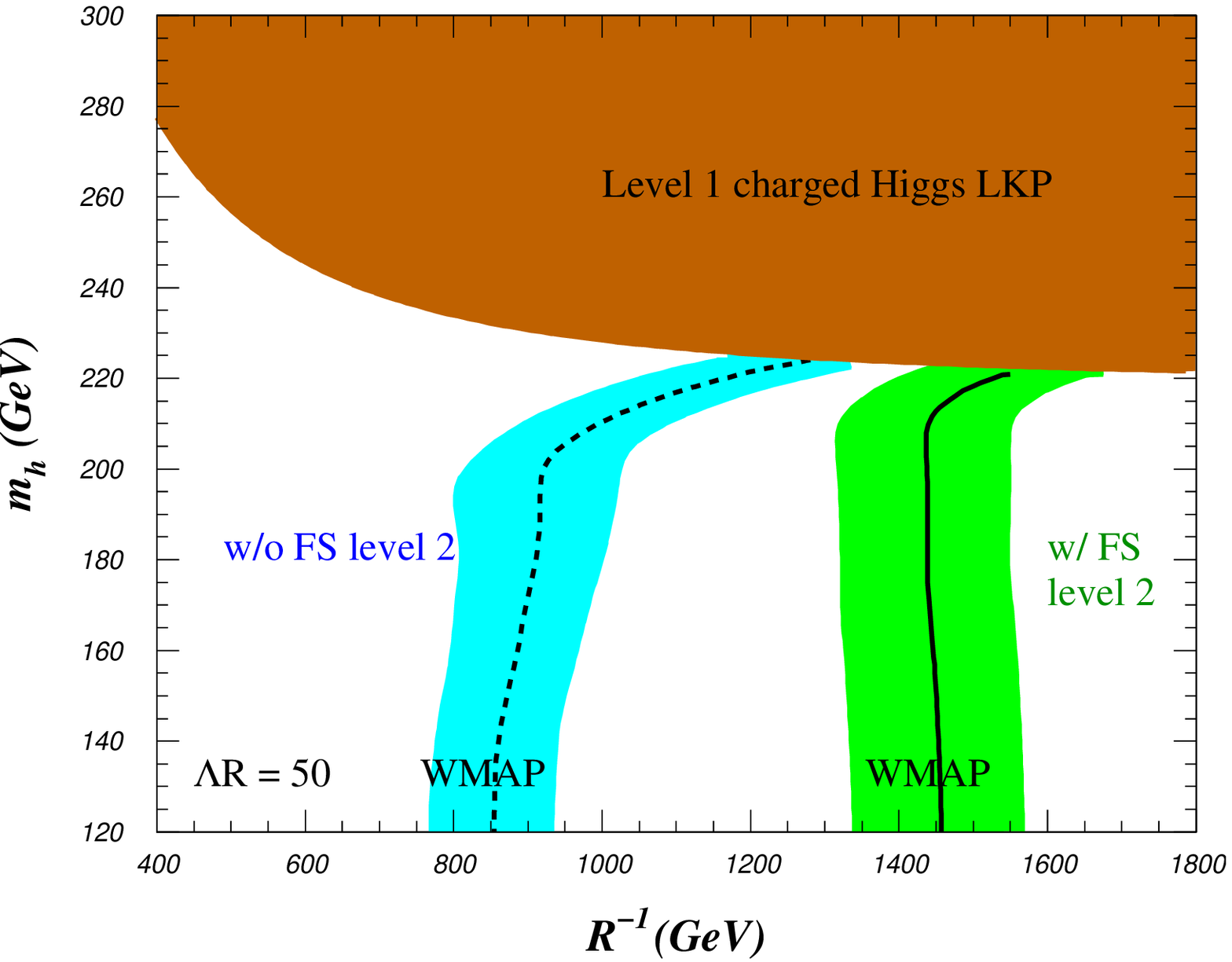}
\vspace{-0.5cm}
\caption{Contour plot of  $\Omega h^2=0.11$ 
in the  $R^{-1}-m_h$ plane for $\Lambda R=20$ (left) and  $\Lambda R=50$ (right). 
The shaded region correspond to the $3\sigma$ WMAP range, $0.0952 < \Omega h^2 <  0.1288$, in the 
case where level 2 KK-particles in the final state are included (dark) or neglected (light grey).  
All coannihilation channels are taken into account. In the region above the full contour the LKP is the charged Higgs. } 
\label{fig:mh}
\end{figure}

The value of the light Higgs mass which enters the loop corrections to the KK Higgs masses, has some impact on the prediction of the relic
density. Increasing the mass from $m_h=120$~GeV increases the contribution of coannihilation channels involving level 1 KK Higgses
thus reducing $\Omega h^2$, see Fig.~\ref{fig:mh}.  The effect is
noticeable only when $m_h>200$~GeV and is of the order of 7\% for  $R^{-1}=1.3$~TeV. 
For Higgs masses around $m_h \approx 220$~GeV, the charged Higgs becomes the LKP~\cite{Matsumoto:2005uh}.

\subsection{Allowing for arbitrary mass splittings}
\label{sec:splitting}

To illustrate the importance of the exact mass splittings between the KK states we generalize our model by allowing 
additional corrections to the fermion masses. In practice this is done by introducing a small shift in $Z_{f}$, Eq.~\ref{eq:zl}.
This procedure guarantees
that gauge invariance is maintained at the tree-level. This also implies that the relative mass shift is applied to both $n=1$ and $n=2$  KK leptons.
First we  illustrate the effect by modifying  only the mass splitting of the KK partners of the lepton singlets, we assume  generation
universality. The free parameters of the model then include in addition to $\Lambda R$, $R$ and $m_h$ of the minimal model, 
five new parameters that affect the masses
of the KK fermions, $Z_{l_R},Z_{l_L}, Z_{Q_L}, Z_{d_R}, Z_{u_R}$. 

\begin{figure}[!ht]
\setlength{\unitlength}{1mm} 
\includegraphics[height=6.cm]{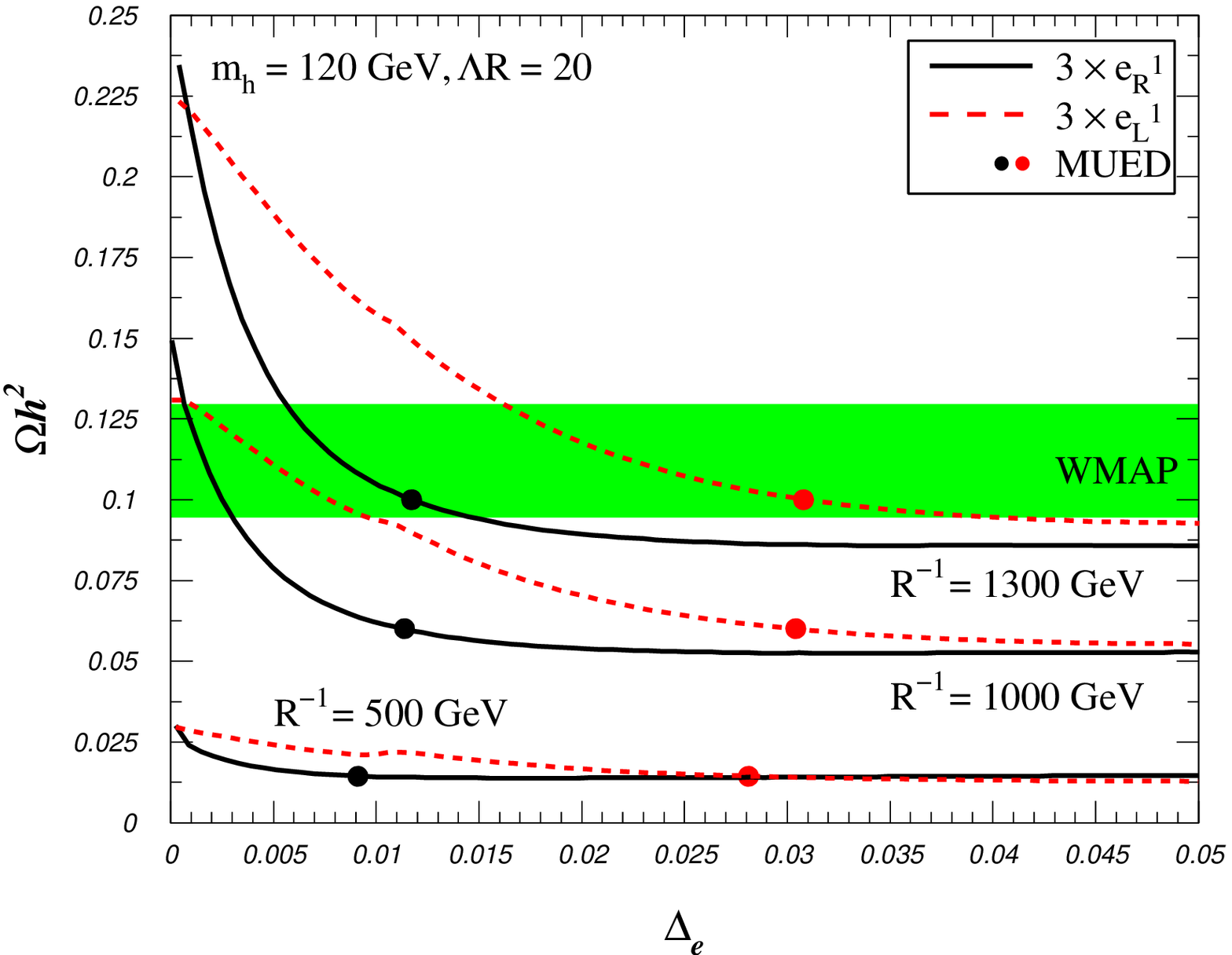}\hspace*{-1mm}
\includegraphics[height=6.cm]{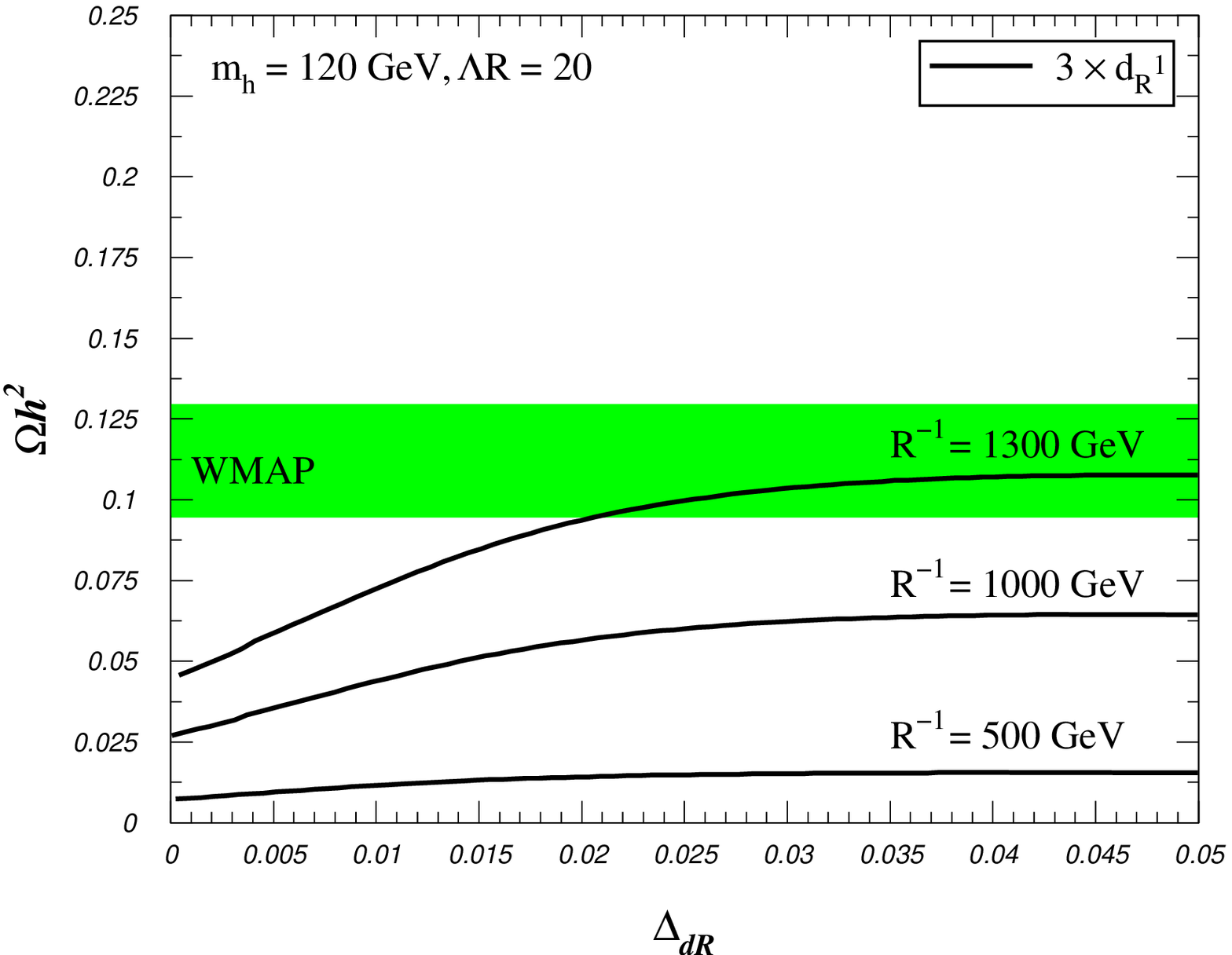}
\vspace{-.1cm}
\caption{$\Omega h^2$ as function of  a) $\Delta_e=(m_{e_R^1}-m_{\gamma^1})/m_{\gamma^1}$ (black) or 
$\Delta_e= (m_{e_L^1}-m_{\gamma^1})/m_{\gamma^1}$ (red-dot)
and b) $\Delta_{dR}=(m_{d_R^1}-m_{\gamma^1})/m_{\gamma^1}$ 
for $R^{-1}=0.5, 1, 1.3$ TeV and $m_h=120$~GeV. The other parameters are as in MUED. Blobs
represent MUED points.  The shaded region corresponds to the $3\sigma$ preferred region obtained by WMAP.} 
\label{fig:deltam}
\end{figure}

We observe, see Fig.~\ref{fig:deltam} for the case of the singlet leptons, that the value of $\Omega h^2$ increases with a smaller mass splitting. This might seem in contradiction 
with the discussion above since we
had argued that the lepton coannihilation had the effect of decreasing $\Omega h^2$,  as
seen from Fig.2, a0 and c0. 
The main effect of a smaller mass splitting is to reduce the contribution of the channel $\sigma(e_R^1 \gamma^1\ra e \gamma^2)$,
indeed  the $e^2_R$ resonance moves very near the threshold for the reaction and so does not contribute significantly to the thermally 
averaged cross section. This effect is more significant than  the increase in the  Boltzmann  factor which can be  
at most 15\%  since in MUED for lepton singlets,  $B_{e_R}=0.86$ for $R^{-1}=1.3$~TeV.
In a sense the relic density is moving towards the value it would have if we had neglected the production of $\gamma^2$ in the 
final state. 
Conversely  an increase  in the mass splitting  leads to a mild decrease in $\Omega h^2$, here the Boltzmann suppression of the 
coannihilation channels is more than compensate by the decrease in the number of degrees of freedom.
Note however that the relic abundance is insensitive to the mass splitting if it is more than 3\%.

We have also examined the effect of the mass splitting with the partners of the left-handed leptons. 
The effect follows the same trend although the influence on $\Omega h^2$ occurs for splittings below 5\%. 
The maximum increase in $\Omega h^2$ is comparable to the one obtained for singlet leptons, see the left frame of Fig.~\ref{fig:deltam}. 
Decreasing the mass of KK quarks on the other hand has the opposite effect as for leptons. 
A smaller mass splitting leads to a lower value for $\Omega h^2$, this is because in this case the factor $B_i$ changes
significantly and QCD processes of the type $q^1q^1\rightarrow q q$ give a large contribution.
To illustrate this we consider the case where we shift the mass
of the KK singlet d-type quarks, see the right frame of Fig.~\ref{fig:deltam}.  
Finally  we  have also considered the implication of mass shifts for the Higgs. The KK-Higgs masses are modified either by increasing 
the light Higgs mass  or by introducing a mass shift via the  parameter $Z_\phi$.
In both cases this can lead to an increase of $\Omega h^2$ around 10\% when the mass difference is a few per-mil.

In summary keeping the mass splitting as a free parameter allows to find scenarios that have 
 $\Omega h^2$ in the range preferred by cosmological measurements for a lower KK scale. 
 For example for  $R^{-1}=1$~TeV, we find $\Omega h^2=0.11$ when $m_{e_R^1}-m_{\gamma^1}=1.7$~GeV or
 $m_{e_L^1}-m_{\gamma^1}=7$~GeV. On the other hand enlarging significantly the mass splittings between the LKP and all level 1 KK particles 
 so as to reduce the contribution of the coannihilation channels would bring us back to the no coannihilation case with a preferred value for the mass of the DM
 candidate around 800GeV, see Fig.~\ref{fig:omega}.

\subsection{Dark matter searches}
\label{sec:direct}

The computation of the LKP-nucleon scattering cross section relevant for direct detection was performed with \micromegas2.4~\cite{Belanger:2008sj}. 
The elastic LKP-nucleon elastic scattering cross section 
depends both on the Higgs exchange as well as on the diagrams with exchange of level 1 KK-quarks. 
The amplitude  is proportional to $1/m_h^2$ for the first contribution and  inversely
proportionnal to the mass difference between the level 1 KK-quarks and the LKP, $1/\Delta^2$  for the second contribution~\cite{Servant:2002hb}.
For the typical mass difference of the MUED scenario, $\Delta= m_{q^1}-m_{\gamma^1}\approx 0.17 m_{\gamma^1}$  the Higgs exchange is dominant. 
When computing the LKP-nucleon cross section we will use the \micromegas~ option that includes  the loop contribution to 
the KK quark exchange diagram. Although this
contribution is computed exactly only for scalar particle exchange,  it gives a  better approximation than the tree-level calculation.
Indeed at tree-level  one can see the effect of the resonance contribution of the $t^1$ diagram when 
$R^{-1}$ is such that $m_q+m_{\gamma^1}=m_{q^1}$. This resonance effect is not physical and is just a sign that including the t-quark contribution by 
taking into account tree-level diagrams together with a coefficient for describing the t-quark content of the nucleon is not a good approximation
~\cite{Drees:1993bu}. 
To compute the elastic scattering cross section we use two sets of quark coefficients in the nucleon,
the default values of \micromegas~, $\sigma_{\pi N}=56, \sigma_0=35$~MeV, as well as values extracted from recent lattice calculations
$\sigma_{\pi N}=47, \sigma_0=42.9$~MeV. The lattice calculations in particular
indicate that the average value for the s-quark operator  $\sigma_s=50$~MeV is lower than expected before~\cite{Giedt:2009mr}.

\begin{figure}[!ht]
\setlength{\unitlength}{1mm} 
\includegraphics[height=6.cm]{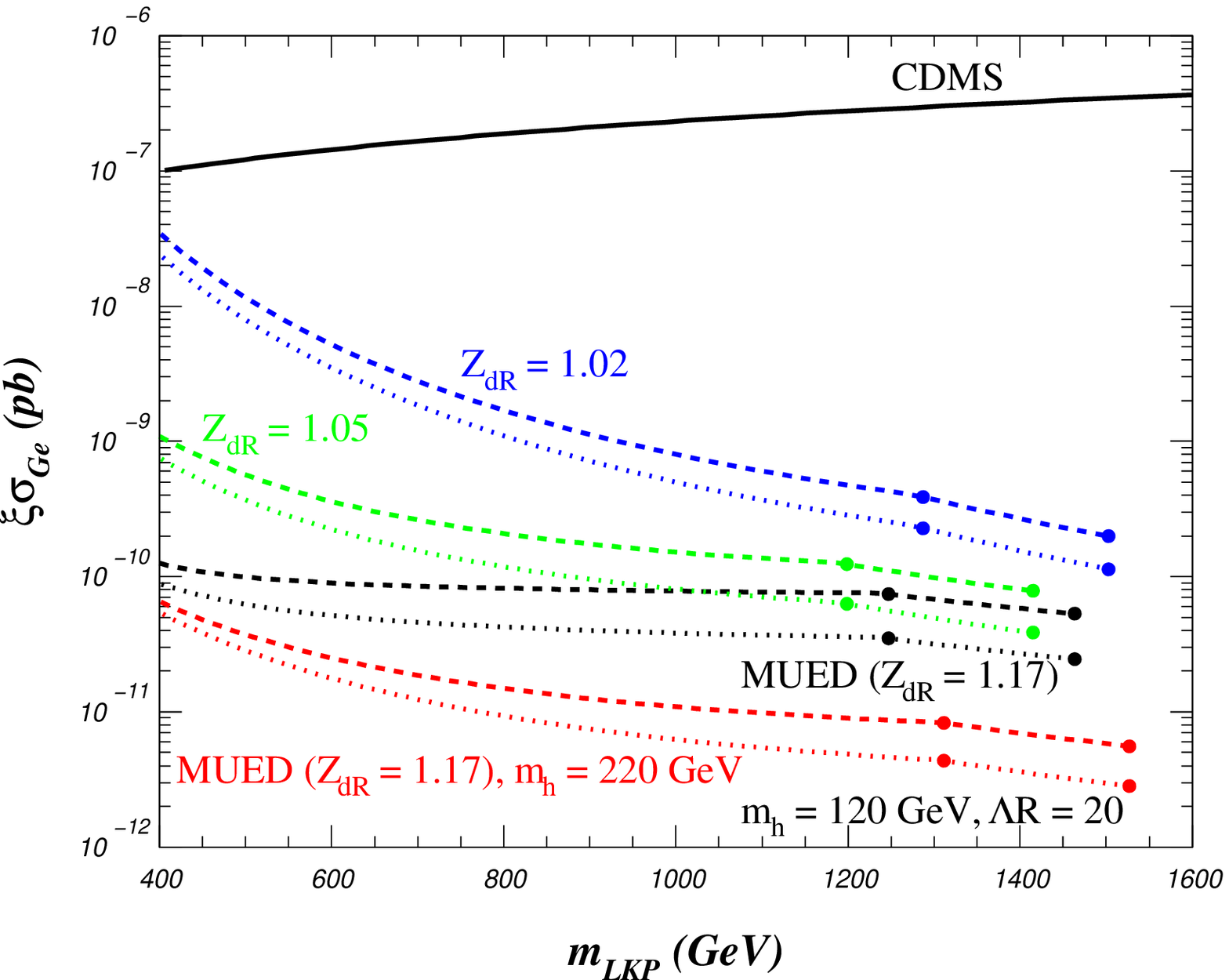}\hspace*{-1mm}
\includegraphics[height=6.cm]{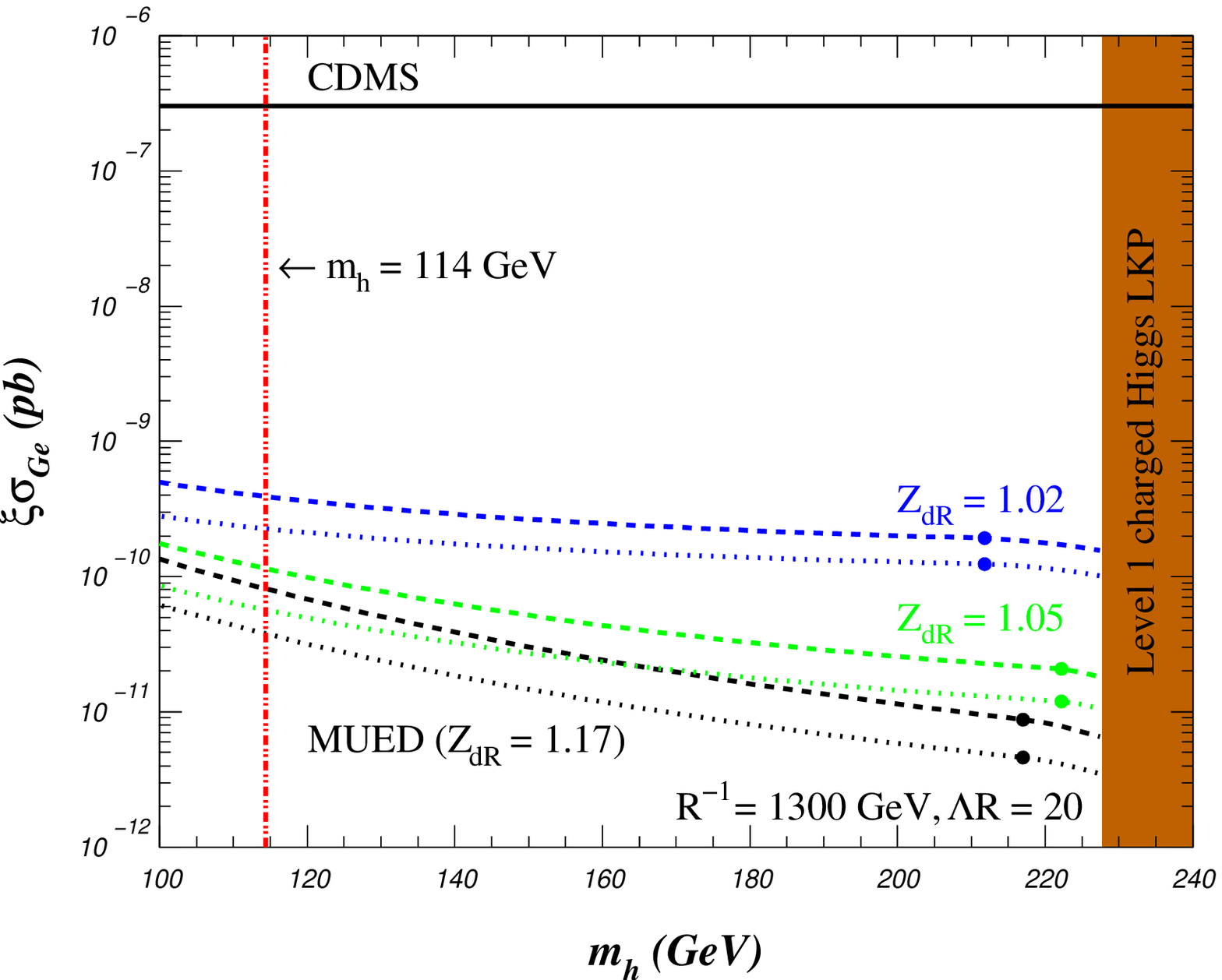}
\vspace{-.3cm}
\caption{Rescaled LKP-nucleon cross section on $Ge^{76}$ vs $m_{LKP}$ for
$m_h=120$ GeV, $\Lambda R=20$ and 2 sets of quark coefficients (
($\sigma_{\pi N},\sigma_0$) = (56 MeV, 35 MeV) (dash) or (47 MeV, 42.9
MeV) (dot) ) and for different values of the mass splitting between
the KK singlet d-quarks and the LKP including the MUED case (left
panel).  The MUED results for $m_h=220$~GeV are also shown.  In each
line the region between the blobs is consistent with the $3\sigma$
WMAP range. Rescaled LKP-nucleon cross section on $Ge^{76}$ vs $m_h$ for
$R^{-1}=1300$ GeV, $\Lambda R=20$ (right).  In each line the region
left of the blob is consistent with the $3\sigma$ WMAP range.}
\label{fig:si}
\end{figure}

In MUED, cross sections are rather low, typically several orders of magnitude below the best limits of CDMS and Xenon. 
We compute the rescaled cross section on point-like nucleus~\cite{Belanger:2010cd}. The rescaling factor takes
into account the fact that $\gamma^1$ does not account for all the dark matter.
It is set to $\xi=\Omega h^2/0.0945$ when DM is underabundant and to $\xi=1$ otherwise.
Although we display results for scattering on $Ge^{76}$, the protons and neutrons contribution do not differ much 
since the contribution from Higgs exchange is dominant, therefore $\sigma_{LKP-Ge}\approx \sigma_{LKP-n} \approx \sigma_{LKP-p}$. 
The rescaled cross section is rather stable around $1\times 10^{-10}$~pb, see Fig.~\ref{fig:si} for the default parameters and is reduced
by roughly a factor 2 using the lattice coefficients. 

In non-minimal models where the mass splittings are treated as free parameters, the LKP-nucleon scattering cross sections can 
increase by orders of magnitude  due to the contribution of the quark exchange diagram which is proportionnal to $1/(m_{q^1}-m_{\gamma^1})^2$~\cite{Servant:2002hb}. 
To illustrate this  we decrease the mass of the right-handed d quarks by treating $Z_{d_R}$ as a free parameter. 
For example for $R^{-1}=1.3$~TeV and $\Delta m\approx 7$~GeV, we find $\sigma \approx  10^{-8}$~pb, about one order of magnitude below 
the limit of CDMS~\cite{Ahmed:2009zw}. 
Of course, introducing such a small mass splitting has also an impact on the relic abundance, the
new coannihilation channels with KK quarks reduce $\Omega h^2$ as discussed in section~\ref{sec:splitting}.
Note that when the mass splitting becomes very small one can see the effect of the resonance contribution of the $b^1$ diagram when 
$R^{-1}$ is such that $m_q+m_{\gamma^1}=m_{q^1}$, this effect
is present even using the loop improved calculation.

For $m_h=220$~GeV, $\sigma^{SI}_{LKP-N}$ drops by almost one order of magntiude in the MUED case, 
this is because the Higgs contribution completely dominates. On the other hand the decrease in the cross section with the larger 
Higgs mass is more modest  in the case where $\Delta_{d_R}$ is small, see Fig.~\ref{fig:si}. This is because 
the Higgs contribution becomes sub-dominant. 
The spin dependent cross section is also typically small in MUED, for example $\sigma^{SD}_{LKP-p}=2.4\times 10^{-7}$pb and 
$\sigma^{SD}_{LKP-p}=2.4\times 10^{-8}$pb for $R^{-1}=1.3$~GeV. These are almost six orders of magnitude below
the current limits ~\cite{Behnke:2010xt,Archambault:2009sm, Angle:2008we}.

\begin{figure}[!ht]
\setlength{\unitlength}{1mm} 
\centering{\includegraphics[height=7.cm]{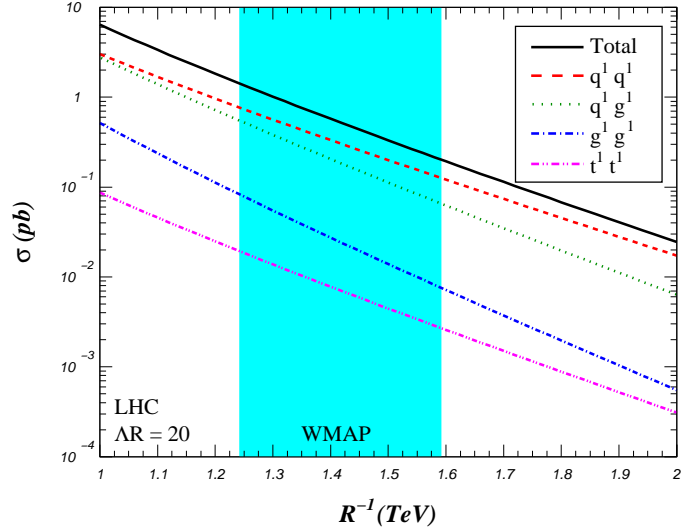}\hspace*{-1mm}}
\vspace{-.2cm}
\caption{ $\sigma(pp\rightarrow q^1 q^1,q^1g^1,g^1g^1,t^1t^1)$ vs  $R^{-1}$ at the LHC for   $\Lambda R=20$ 
using the CTEQ6M PDF's. The K-factor is not included. The shaded region corresponds to the $3\sigma$ preferred 
region obtained by WMAP.}
\label{fig:lhc}
\end{figure}

For completeness we have also computed typical cross sections for KK-particles production at LHC with $\sqrt{s}=14$~TeV .
The largest  cross sections  are obtained for coloured particles. The dominant process are $q^1q^1$ pairs followed by $q^1g^1$, both with 
cross sections in the range  ${\cal O}(.1-1)$~pb for the DM preferred mass scale, see  Fig.~\ref{fig:lhc}.
These results are in agreement with previous calculations ~\cite{Cheng:2002ab,Macesanu:2002db,Macesanu:2005jx,Bhattacharyya:2009br} 
and are slightly suppressed as compared with the ones obtained using only tree-level KK masses.

\section{Conclusion}

We have performed a complete computation of the relic density of dark matter
in the minimal UED model including all effects of level 2 particle in the intermediate and final state. 
We have shown that the production of $\gamma^2$ in the final state reduces significantly the relic density thus shifting
the preferred region from 500-600~GeV to well above the TeV scale (more precisely around 1.3TeV). 
The slight  tension between the electroweak precision observables which favour $R^{-1}>600$~GeV~\cite{Appelquist:2002wb,Gogoladze:2006br} for a
Higgs mass of 115~GeV
and the dark matter observables is thus released. Indeed for $R^{-1}=1.3$~TeV  the contributions  to the $S$ and $T$ parameters are suppressed, 
with in particular $T\approx  10^{-4}$. On the other hand the higher scale means that it will be harder to probe these scenarios at 
 LHC  as well as in direct detection. Furthermore because of the important contribution of the coannihilation channels, the typical value of $\langle
 \sigma v \rangle$ relevant for indirect cross section signals is suppressed relative to the typical cross section 
 expected for models that give $\Omega h^2=0.1$.
Generalizing the model to allow for arbitrary mass shifts in the KK spectrum,
we have shown that one could again increase the relic density so that agreement with WMAP was recovered for a LKP around the TeV scale  in the case where the
lepton NLKP were almost degenerate with the LKP. In this case the direct detection cross section could be strongly enhanced. 

We have also shown that these results  not only depend sensitively on the mass difference between the level 1 particle and the LKP but also on the  
precise mass of the level 2 particles that can be exchanged in s-channel in annihilation or coannihilation processes. 
In that sense making a precise theoretical prediction of the relic abundance of DM based on collider observables~\cite{Allanach:2004xn}
in the event of the observation of KK-particles at the LHC is expected to  be
extremely challenging. Indeed  it  would require not only the measurement of the mass and couplings of the LKP but also 
a precise determination, in some cases better than the percent level, of the masses of level 2 particles. 
Such precision is not within the reach of the LHC especially for particles that are well above 1 TeV.

\section{Acknowledgements}
We thank A. Semenov for his help with LanHEP. We also thank A. Belyaev and F. Boudjema for useful discussions.
This work was supported by HEPTOOLS under contract MRTN-CT-2006-035505.
This work was also supported in part by the GDRI-ACPP of CNRS and by the French 
ANR project {\tt ToolsDMColl}, BLAN07-2-194882.
The work of AP was supported by the Russian foundation for Basic Research, 
grant  RFBR-08-02-92499-a, RFBR-10-02-01443-a.

%\bibliography{ued}{}
\providecommand{\href}[2]{#2}\begingroup\raggedright\endgroup

\end{document}